\documentclass{article}
\usepackage{times}  
\usepackage{helvet}  
\usepackage{courier}  
\usepackage[hyphens]{url}  
\usepackage{graphicx} 
\usepackage[numbers]{natbib}  
\usepackage{caption}

\usepackage{arxiv}
\usepackage{amsmath}
\usepackage{graphicx}
\usepackage{xcolor}
\usepackage{amssymb}
\usepackage{yfonts}
\usepackage{hyperref}
\usepackage{textalpha}

\title{A novel RNA pseudouridine site prediction model using utility kernel and data driven parameters}

\author{
  Sourabh Patil\\
  Dept. of Biological Sciences, BITS Pilani K K Birla Goa Campus, Goa, India\\
  \texttt{f20190572@pilani.bits-pilani.ac.in,}
\And
Archana Mathur\\
Dept. of Information Science and Engineering, Nitte Meenakshi Institute of Technology, India\\
\texttt{archana.mathur@nmit.ac.in}
\And
Raviprasad Aduri\\
Dept. of Biological Sciences, BITS Pilani K K Birla Goa Campus, Goa, India \\
\texttt{aduri@goa.bits-pilani.ac.in}
\And
Snehanshu Saha\\
Dept. of CSIS and APPCAIR,  BITS Pilani Goa, India\\
HappyMonk AI, Bangalore, India\\
\texttt{snehanshus@goa.bits-pilani.ac.in}
}

\begin{document}
%


\maketitle              
\begin{abstract}
RNA protein Interactions (RPIs) play an important role in biological systems. Recently, we have enumerated the RPIs at residue level and have elucidated the minimum structural unit (MSU) in these interactions to be a stretch of five residues (Nucleotides/amino acids). Pseudouridine is the most frequent modification in RNA. The conversion of uridine to pseudouridine involves interactions between pseudourdine synthase and RNA. The existing models to predict the pseudouridine sites in a given RNA sequence mainly depend on user defined features such as mono and dinucleotide composition/propensities of RNA sequences. Predicting pseudouridine sites is a non-linear classification problem with limited data points. Deep Learning models are efficient discriminators when the data set size is reasonably large and fails when there is paucity in data ($<1000$ samples). To mitigate this problem, we propose a Support Vector Machine (SVM) Kernel based on utility theory from Economics, and using data driven parameters (i.e. MSU) as features. For this purpose, we have used position-specific tri/quad/pentanucleotide composition/propensity (PSPC/PSPP) besides nucleotide and dineculeotide composition as features. SVMs are known to work well in small data regime and kernels in SVM are designed to classify non-linear data. The proposed model outperforms the existing state of the art models significantly ($10\%-15\%$ on average).  

\end{abstract}
\newpage
\section{Introduction}

Pseudouridine ($\psi$) is a modified nitrogenous base commonly found in RNA. It is formed by the isomerization of uridine, a natural RNA building block, to provide structural and functional diversity. $\psi$ is present in transfer RNA (tRNA), ribosomal RNA (rRNA), and small nuclear RNA (snRNA) among other types of RNA \cite{charette2000pseudouridine}. $\psi$ sites are highly conserved across different species, indicating their functional significance. The discovery and characterization of $\psi$ sites have been facilitated by advances in high-throughput sequencing technologies and computational approaches. These techniques enable researchers to map and identify $\psi$ sites on a global scale, providing insights into their distribution and functional roles across various RNA molecules. Understanding the role of $\psi$ sites in RNA biology continues to be an active area of research with potential implications for various fields, including molecular biology, genetics, and medicine. 
Finding $\psi$ sites in RNA is desirable for several reasons:

\begin{itemize}
    \item Understanding RNA modification: Understanding RNA post-transcriptional modifications enables researchers to unravel the regulatory mechanisms that control RNA processing, stability, and function. 
    
    \item RNA structure and function: 
    Role of $\psi$ modifications contribute to RNA folding, stability, and interactions with other molecules. This knowledge helps elucidate the molecular mechanisms underlying RNA-mediated processes such as translation, splicing, and RNA-protein interactions. 
    \item Biomarker discovery: 
    Discovering $\psi$ sites can lead to the identification of potential RNA biomarkers associated with specific physiological conditions or pathologies. 
    \item Evolutionary conservation:
    Comparative studies of $\psi$ sites can shed light on the evolutionary relationships between species and help decipher the functional consequences of $\psi$ modifications in different biological contexts. 
\item RNA-targeted therapies:
    The discovery of $\psi$ sites can aid the development of RNA-targeted therapies. RNA modifications, including pseudouridine, can influence the efficacy and specificity of RNA-targeting drugs, such as antisense oligonucleotides or mRNA-based vaccines.
\end{itemize}

\noindent Recently, $\psi$ site prediction using AI tools has gained prominence owing to the above mentioned reasons.The performance of these methods are far from ideal given the advancements in both AI tools and biological data representation. Pseudouridylation of RNA involves the structural context of both the protein and RNA. One major contributing factor for a functional RNA protein interaction is the structural context of the two partners \cite{lunde2007rna,geisler2013rna}. All the existing methods are agnostic to the structural features of RNA while representing the data. Our recent work on predicting RNA protein interactions (RPI) revealed the presence of a minimum structural unit (MSU) in RPI \cite{xRPI}. This MSU consists of a stretch of five nucleotides/amino acids in RNA/protein for a stable interaction to occur between the two partners. Based on this knowledge, we have used position specific tri, quad, and penta nucleotide propensities besides the mono and di nucleotide compositions as features and utility kernel, here we propose an SVM model for predicting $\psi$ sites in RNA.

\section{Related Work/Contributions:} 
There are two existing methodologies viz iRNA-PseU \cite{Chen2016iRNAPseUIR} and PseUI \cite{He2018PseUIPS}. 
The iRNA-PseU had Nucleotide density and pseudo-nucleotide composition (PseKNC) as features. 

On the other hand, the PseUI had five different features in their work - Nucleotide composition, Dinucleotide composition, Position-specific nucleotide propensity (PSNP), Position-specific dinucleotide propensity (PSDP), and PseKNC. They used the RBF kernel of the Support vector machine (SVM) to solve the challenge of predicting the RNA pseudouridine site.

The best performance these methods could achieve is an accuracy of around 70 \%\ auguring for improving the existing state of the art. To improve any ML based model, one needs to consider the feature representation and the ML architecture. Herein, we have proposed RNA features based on the knowledge derived from our RPI work \cite{xRPI} and by taking a cue from the utility kernel applied in the domain of economics. We refer to our model as 'PSe-MA'. 
\subsection{Our Contributions}
\begin{enumerate}
     \item For any prediction model that involves structural context of a given biomolecule, as is the case with $\psi$ site prediction in RNA, we propose using data driven parameters as well as the minimum structural unit (MSU) of the given biomolecule as one of the features. 
    \item We present the insight into the Small Sample Size ($S^3$) classification problem as the key driving point behind our data-driven investigation as we noticed that the baseline models that used SVM RBF kernels didn't provide any justification! 
    \item The insight (refer figure \ref{fig:shap1}) leads to the discovery of a novel SVM Kernel. We propose a parameterized utility kernel for the task presented in this work.
    \item We empirically validated the discrimination ability of the kernel and showed that it outperforms the previous baseline and other ML algorithms by a significant margin.
    \item Further, We present the Mathematical and computational foundation of the novel Kernel which explains the superior performance.   
\end{enumerate}
\section{Methods}
This study used three benchmark datasets for training: H990, S628, and M944. A datailed description of each of them can be found in the 'Data' section below. These are the same datasets used by iRNA-PseU and PseUI methods. These datasets are curated from the experimentally found $\psi$  sites from RMBase for H. sapiens, M. musculus, and S. cerevisiae. The negative dataset is made of RNA sequences that were experimentally confirmed to not have $\psi$ sites. For more detailed information on how these datasets were constructed, refer to \cite{Chen2016iRNAPseUIR}.
Each of the datasets has equal number of positive and negative samples (i.e. RNA sequences with and without a pseudouridylation site). For example, H990 contains 495 sequences with a $\psi$ site and 495 sequences with uridine that doesn't get pseudouridylated. 
We denote each RNA segment as $R_{\epsilon}$(U), in these datasets as: $R_{\epsilon}$(U)=$N_{-\epsilon} N_{-(\epsilon-1)}.....N_{-1}UN_{1}....N_{(\epsilon-1)}N_{\epsilon}$
where the center U represents ‘uridine’, $N_\epsilon$ represents
the $\epsilon$-th upstream nucleotide from the central uridine
and N+$\epsilon$ represents the $\epsilon$-th downstream nucleotide.
The RNA samples in H990 and M944 are composed of 21 nucleotides each, while those in S628 consist of 31 nucleotides. Hence, the $\epsilon$ is 10, and the RNA segment length is 2 × 10 + 1 for the H900 and M944 datasets. The value of $\epsilon$ is 15, and the RNA segment length is 2 × 15 + 1 for the S628 dataset.
In addition to the training datasets, we have used two independent testing datasets, namely H200 and S200, provided by Chen et al. \cite{Chen2016iRNAPseUIR} for validation.

\subsection{Feature representation of RNA samples}
Encoding an RNA sequence as a feature vector with highly discriminative features is one of the main challenges in building a predictor based on machine learning. 
This is due to the fact that all currently available machine learning methods can only handle vectors with identical lengths for all sequence samples. A vector formed in a discrete paradigm, however, can lose its sequence-pattern details. 
Here, we propose five different features to represent the RNA sequences: nucleotide composition (NC); dinucleotide composition (DC); position-specific trinucleotide propensity (PSTP); position-specific quadnucleotide propensity (PSQP); and position-specific pentanucleotide propensity (PSPP, which is also the MSU for RNA) features. The NC and DC are calculated using the the Pse-in-One server \cite{Liu2015PseinOneAW}. Below is a brief description of the features used in the current study.
\begin{itemize}
    \item Nucleotide composition (NC):
  NC refers to the frequencies of each nucleotide (A, U, G, and C) within a sequence or a specific region of interest.
    \item Dinucleotide composition (DC):
   DC, also known as di-nucleotide frequency or di-nucleotide content, is a feature commonly used in computational biology to take advantage of the context dependence of the nucleotides (i.e. the presence/absence of a specific nucleotide either on the 5\'\ or the 3\'\ side of the given nucleotide.
    \item Position-specific trinucleotide propensity (PSTP):
   PSTP provides a measure of the effect of nearest neighbors in the axiom of sequence to structure to function. Instead of accounting for either the upstream or downstream nucleotide identity as done in DC, herein, one can incorporate the information of both the up and downstream nucleotide identity for a given nucleotide at a given position in the sequence. This provides a basic measure of the structural context of each of the nucleotides in the RNA sequence.
    \item Position-specific quadnucleotide propensity (PSQP):
   PSQP expands upon the PSTP in terms of the number of nearest neighbors being considered for the feature generation. The tetra nucleotide composition is considered here based on the knowledge, from the literature and our work on RPIs involved in alternative splicing, that most of the RNA sequence motifs that recognise/involved in the RPIs are of four nucleotides in length \cite{AS-TNF}.
    \item Position-specific pentanucleotide propensity (PSPP):
   The final feature we have constructed for this study is the PSPP. In the case of biomolecular interactions, for example RPI, the structure dictates the interaction. In case of RNA, the minimum structural unit, the hairpin structure, is of five nucleotides. Our previous studies on RPI enumeration also revealed the MSU to be of five nucleotides. Hence, we have constructed the PSPP feature to keep the structure information of the given RNA sequences intact. 
\end{itemize}
\vspace*{-6mm}
\section{RNA pseudouridine site prediction: The ML Challenge}
\textbf{Data:} As shown in the tables (see supplementary file, section D tables 7-8), three datasets were used in this study corresponding to species \textit{H.sapiens}, \textit{S.cerevisiae} and \textit{M.musculus} respectively. It is a binary balanced classification problem with an equal number of positive and negative samples. Class 1 corresponds to the RNA sequences which have a pseudouridine site present in them and Class 0 corresponds otherwise. \textit{Therefore, RNA pseudouridine site prediction is a supervised machine learning (classification) problem.}
In addition to the training datasets, Chen et al. \cite{Chen2016iRNAPseUIR} provided two independent testing datasets for H. sapiens and S. cerevisiae, namely H200 and S200, but not for M. musculus.
 
\begin{figure*}[h]
\centering
\includegraphics[width=.30\textwidth]{ 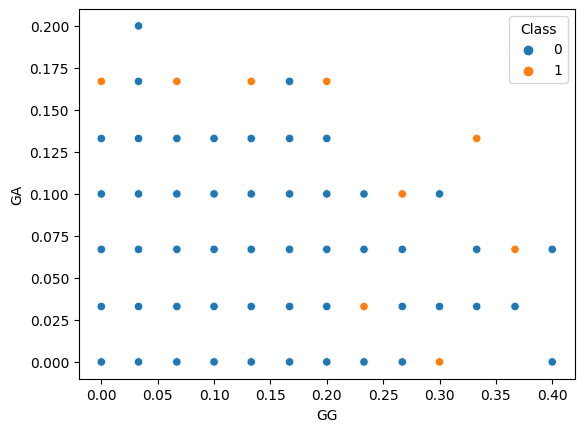}
\includegraphics[width=.30\textwidth]{ 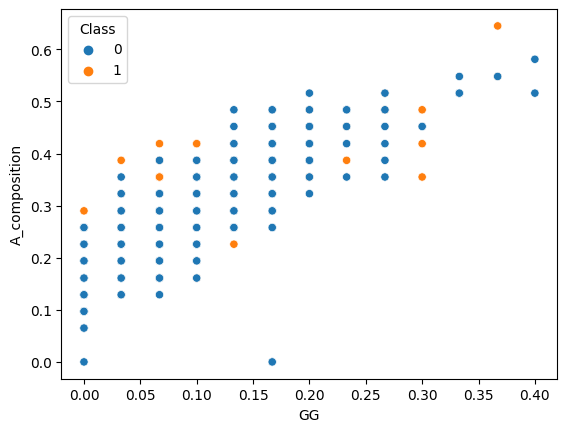}
\includegraphics[width=.30\textwidth]{ 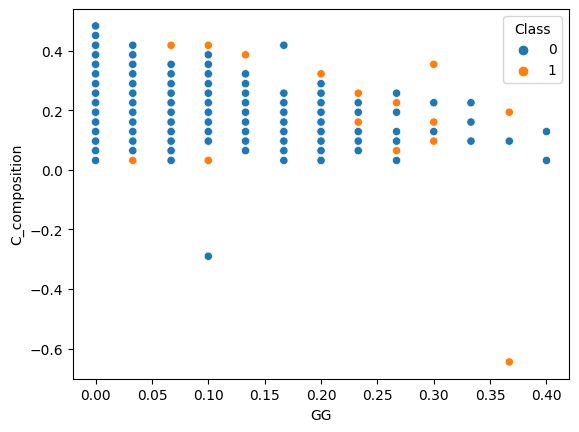}
\caption{Feature-feature plots of the dataset (RNA pseudouridine site prediction): non-linear decision surface is clearly visible i.e. linear model/surface can't separate (shatter) the red points from the blue ones.}
\label{fig:shap1}
\end{figure*}

Deep Learning models fail to perform well when the sample size is small \cite{Rajput2023EvaluationOA} \cite{Keshari2020UnravellingSS}, particularly when the samples are less than a thousand. Additionally, if the data is non-linear in nature, the classification tasks become even more challenging. Deep Learning models efficiently discriminate non-linear decision surfaces if and only if the models are trained on sufficiently large data as they can generate new features from the corpus of data and find separating hyperplanes. The notion of being able to approximate data patterns via the Universal approximation theorem \cite{Saha2020EvolutionON} also rests on this assumption. Therefore, the choice of a suitable Machine Learning (ML) model is restricted to the ones that are not data-hungry! Support Vector Machines (SVM) is one such robust ML model grounded on sound statistical learning theory, effectively operating in the small sample regime (known as small sample size or $S^3$ problems). They are known to perform well on linear (linear-SVM) as well as non-linearly separable datasets (SVM kernels) (refer figure \ref{fig:shap1}). The idea behind SVM kernels is to transform the training data into higher dimensional space and find an optimal hyperplane that separates the data into distinct classes. In the language of computational learning theory, these attributes are known as \textit{Universal Approximation power, Generalization ability, Error bounds, and VC Dimensions}.
\subsection{Solution: Our SVM kernel Model} 
\noindent 
Based on the property of inner product, which can be extended to a Hilbert Space produces  $ \langle x, y \rangle = x_1 y_1 + x_2 y_2 $, we adapted a class of utility functions, novel kernel, to the properties of the inner product. 
In general, a direct (or indirect) utility function with the display of properties such as reflexivity, monotonicity can be classified within the broad range of von Neumann Morgenstern (sequence of) utility functions \cite{Chichilnisky1985VonN} with point-wise convergence or with almost anywhere convergence. 
This paper utilizes the inner products in lieu of utility functions preserving existing characteristics including the uniqueness of optimal choices supported by budget constraints that are also inner products in the price and commodity space. Suppose, the budget set facing a consumer is given by:
 $Bp_w =  \{x | px \leq w \}$
where, $w  =$ income or wealth; the budget frontier (line or hyperplane) is given by $\overline{B} = \{x|px = w \}.$ If $p$ is orthogonal to a budget frontier $\overline{B}$, then if, $x,y \in \overline{B}$ , then the inner product is $p(y-x)$, where, $p(y-x)= py-px = w-w = 0$. We leverage these ideas to construct a Utility kernel for classification. Von Neumann-Morgenstern utility functions the degree of risk aversion (in the case of
Hyperbolic Absolute Risk Aversion (HARA) \cite{merton1975optimum}, it is the risk tolerance instead) can classify elements according to distinct risk-taking abilities, ex-post.
In all unconstrained cases, the utility function itself may operate as the classifier. Indeed, in this case, the utility functions are raised to power with the help of a risk aversion parameter, which can well be a (non-zero) integer used as a scaling factor, or restricted between zero and one, in order to obtain a classification. This sets the ground for our Utility Kernel,\\
\noindent \textbf{The Utility Kernel:} We propose a Utility kernel, expressed as:
\begin{equation}
    K(x_i,x_j) = k_0 +k_1 <x_1, x_2>^\alpha
    \label{eqn:1}
\end{equation}
 where $k_0, k_1$ are tunable parameters and $\alpha$ is the degree of the proposed kernel.


\subsection{The Mathematical foundation of Utility kernel: Generalization ability, Error bounds and VC Dimensions}\label{sec:gen_ability}
We state some important theoretical findings related to the utility kernel. The proofs are available on request. These results provide the foundation for the utility kernel (UK) in the sense that they state (1) why the utility inner product qualifies as an SVM kernel (Mercer's Theorem), (2) how good is the approximation capability of the UK on the trained samples (UK as universal approximator), (3) performance on the unseen test data (Generalization ability, Error bounds) and (4) the ability of UK to shatter (classify) the largest set of points i.e. it is a measure of the capacity of a hypothesis set to fit different data sets or the measure of the complexity of the UK model and its fitting ability on different data sets (VC dimension). A Kernel must satisfy Mercer's theorem as follows.\\
\textbf{Mercer's Theorem:}  If $K$ is a continuous symmetric function such that the integral operator $L_K:L^2 \rightarrow L^2$, defined as $L_{K} g(x) = \int K(x,y) g(y)dy$ is positive, then for all functions $g \in L^2$, the  condition $\int \int K(x,y) g(x) g(y) dx dy \geq 0$ stands true.\\
\textbf{Utility kernel as universal approximator:} \\
Theorem: A function $\mathfrak{F}$ can be called a universal approximator if, for a compact set $C\subset \Re ^m$, a continuous function h can be found such that the maximum norm $\left | f(x)-h(x) \right |\leq \epsilon $ for $f \in \mathfrak{F}$ and $\epsilon > 0$. Let there be m-dimensional non-linearly separable patterns $\left (x_1, y_1  \right )\cdots (x_n,y_n)$. Let $\phi :\Re ^m\rightarrow X$ be a non-linear mapping, where $X$ is a higher dimensional Hilbert space, $w$ be the weight vector, $b$ be the bias, '.' be the dot product, $\mathbf{H}$ be the Heaviside function, $\mathbf{H}:\Re \rightarrow \left \{ -1,1 \right \}$. Assuming that we use utility kernel to train the samples, after solving the Kuhn-Tucker conditions and the dual optimization problem, we get a classifier of the form,
\begin{equation*}
    x = \textbf{H}\left ( \sum_{j}^{}\alpha_j y_y k\left ( x_j,x \right )+b \right )
\end{equation*}
where $\alpha_j$ is the Langrange multiplier, $x,x_j \in \Re$. Let $C_g^H(k)$ be the general class of functions with arbitrary $\alpha_j$, $x_j$ and b, k being the utility kernel.\\
\textbf{Preposition: 4.1} The dimension of feature space of the utility kernel $k_u(x_1,x_2)$ is\\
\begin{equation*}
    dim(k_u) = \left\{\begin{matrix}
\binom{n+d-1}{d}+1 & if d\neq n\\ 
\frac{1}{2}\binom{n+d}{d}+1 & otherwise
\end{matrix}\right.
\end{equation*}
\noindent \textbf{Theorem:} $\bigcup_{u=1}^{\infty }C^\textbf{H}(k_u) $ possess the universal approximation capabilities if the kernel $k_u$ is defined as 
\begin{equation}
    k_u(x_1,x_2)=k_0+k_1 \left ( x_1 \cdot x_2 \right )^u
    \label{eq:kernel}
\end{equation}
\textbf{The VC dimension of the Utility Kernel:} \\
Valiant \cite{Valiant1984ATO} proposed that any learning algorithm can find a hypothesis or rule that approximates the best possible rule with some degree of probability.\\
\textbf{Theorem:} Assume training examples $S = \left \{ x_{1},x_{2} ...x_{l} \right \} \in \Re ^{d}$ when transformed by function $\phi$, the space of the transformed features is bounded by smallest hyper-sphere with radius $R$ and center $C$. The VC dimension $VC_{K}$ for the Utility kernel $K(,)$ is bounded by the following inequality -
\begin{equation*}
    VC_{K} \leq min\left ( \left [ \frac{R^{2}w^{2}}{4} \right ],d \right ) + 1
\end{equation*}
where $    R^{2} = K(x_p,x_p) - 2\sum_{i=1}^{l}\beta _i K(x_i,x_p) + \sum_{j=1}^{l}\sum_{k=1}^{l}\beta _j\beta _k K(x_j,x_k)$ and \\ $
    <w,w> = \sum_{i=1}^{l} \sum_{j=1}^{l}\alpha_{i} \alpha _{j}y_{i}y_{j} K(x_i, x_j)$, $\alpha_i,\alpha_j,\beta _i,\beta _j,\beta _k $ are the Lagrangian multipliers.  $\phi x_p$ is the farthest training sample lying on the circumference of the hyper-sphere.The values are essentially data-dependent, as we compute inner products of pairwise samples from the training set. We computed the kernel parameters on the curated data. 

\section{Experimental Settings } 
We conducted our experiments on an Intel(R) Core(TM) i3-8145U laptop with 8GB RAM. Python 3.10 was used to build the ML models and the Utility Kernel was coded from scratch. The grid search was implemented to find the optimal parameters of $k_0$, $k_1$, and $\alpha$. All experiments were run three times and mean and standard deviation of accuracy and other metrics were reported. Along with accuracy, sensitivity, and specificity, the Matthews Correlation Coefficient (MCC), and Area Under the Curve (AUC) are reported for each experimental run.
 
\noindent\textbf{Validation techniques:} Cross-validation (CV) techniques used in the ML domain ensure that the trained model performs well on unseen data. The main idea behind CV is to prevent the model from doing overly well during training but performing miserably when tested in the new data (overfitting). The most popular choices in CV techniques include jackknife, 66\%-33\% train-test split, and k-fold cross-validation. Jackknife (aka Leave-one-out CV) takes the entire dataset for training but leaves one sample for testing. The train-test batches are created to train the model and the performance is averaged in the end. This ensures a low bias in the results but the presence of an outlier may lead to high variance. The 66\%-33\% train-test split is the standard way of CV where 66\% of data is used for training and 33\% of data is used for reporting the performance of the model. In K-fold CV, the data is divided into K-folds. The $k-1$ folds are used for training and the $k^{th}$ fold is tested, and the process is repeated $k$ times, each time with a different subset for train and test. 

\noindent\textbf{Robustness Settings:} We run our method 3 times and report the average and standard deviation of the performance metrics. This is done to ensure the reproducibility of the results under identical settings. We expect minimum statistical variance which should be the desired outcome. Additionally, we run Bonferroni comparisons \cite{Armstrong2014WhenTU} which count the number of times a proposed method turns out to be superior to the baselines and other methods. For example, in order to claim a method as SOTA, it must display superior performance in the majority of the runs (at least 3 out of 5, 6 out of 10 for instance). We computed the Bonferroni Statistic for our method PSe-MA (on Utility Kernel). The BC value for PSe-MA was 3 (out of 3) in all the datasets.  
 
\noindent \textbf{Performance Evaluation metrics:}
In classical ML, a confusion matrix adds a lot of visibility while solving the classification problem, whether binary or multi-class. While testing the model, the confusion matrix presents a count of actual and predicted labels of the classifier. It outputs "TN", which is the number of actual 'negative' samples predicted as negative, and "TP" indicates the number of actual 'positive' samples predicted as positive. "FN" implies positive samples wrongly predicted as negative and "FP" indicates negative samples wrongly predicted as positive. We have used 5 types of evaluation metrics viz. sensitivity, specificity, accuracy, MCC, and AUC (Area under the curve). Sensitivity is the ratio of true positives and the actual positives in the data, given as, $Sensitivity= \frac{TP}{TP+FN}$. Specificity is the ratio of true negatives to the total negatives of the model, represented as $Specificity= \frac{TN}{TN+FP}$. Accuracy is the measure of correct predictions from the total predictions. The formula is $Accuracy= \frac{TP+TN}{TP+TN+FP+FN}$. MCC is the correlation between the actual and predicted values of the classifier and is defined as, 
$MCC= \frac{TP*TN-FP*FN}{\sqrt{(TP+FP)(TP+FN)(TN+FP)(TN+FN)}}$. AUC is obtained from the ROC Curve 
by plotting sensitivity against 1 - specificity. AUC is desired to be close to 1 for superior classification performance.
 
\noindent \textbf{Generalization settings:} The generalization setting is designed for a specific test in mind. It is important to know the efficacy of a classifier trained on one data set and be tested on a completely new data set with identical features as the previously generated data. For this 'blind' validation, there are two separate datasets for \textit{H. sapiens} and \textit{S. cerevisae} each of them have 200 samples (100 positive and 100 negative) More details about these datasets can be found in the supplementary text section D. The setting is different from using a train-test-validate split used traditionally. Needless to mention, Jackknife validation doesn't apply in this setting. 
\begin{table}[t]
    \centering
    \scalebox{0.60}{
 \begin{tabular}{ |p{1.5cm}|p{3cm}|p{2.0cm}|p{2.0cm}|p{2.0cm}|p{2.0cm}|p{1.8cm}|}
\hline
\multicolumn{7}{|c|}{PSe-MA (Utility kernel) using jack-knife CV with PseUI and iRNA-PseU models} \\
\hline
Training datasets& Model used &Sensitivity&Specificity&Accuracy& MCC&AUC\\
\hline
H990   & iRNA-PseU (rbf) &61.01& 59.80& 60.40& 0.21&0.64\\
& iRNA-PseU (linear) & 63.56& 60.21 &61.01  & 0.31&0.34 \\
& \textbf{iRNA-PseU (UK)} &\textbf{66.21±0.02} &\textbf{59.98±0.09} &\textbf{ 61.92±0.05}& \textbf{0.34±0.01} &\textbf{0.55±0.008}\\
& PseUI (rbf) &64.85& 63.64& 64.24& 0.28&0.68\\
&PseUI (linear) &66.78& 65.87&  67.89& 0.35 &0.56  \\
& \textbf{PseUI (UK)} &\textbf{70.76±0.06 }& \textbf{69.54±0.009 }& \textbf{71.33±0.01} &\textbf{ 0.46±0.05} &\textbf{ 0.50±0.04}\\
&  {PSe-MA (rbf)} & {70.71}& {81.21}& {75.96} & {0.52}& {0.76}\\ 
&  {PSe-MA (linear)} & {78.38}& {84.44}& {81.41} & {0.63}& {0.81}\\
& \textbf{PSe-MA (UK)} &\textbf{90.04±0.06}&\textbf{91.14±0.03}&\textbf{90.60±0.09} &\textbf{0.84±0.01}&\textbf{0.90±0.02} \\ 
\hline   
S628 &iRNA-PseU (rbf) &64.65 &64.33 &64.49 &0.29& 0.81\\
& iRNA-PseU (linear) &65.64 &67.66  &67.64  &0.39 &0.63 \\
&\textbf{iRNA-PseU (UK)} &\textbf{68.55±0.001}& \textbf{69.33±0.02}& \textbf{68.61±0.09}& \textbf{0.51±0.06}& \textbf{0.89±0.08}\\
& PseUI (rbf)  &62.10 &71.02 &66.56& 0.33& 0.69\\
&PseUI (linear) &63.56 & 68.92 & 68.71 &0.23  &0.56 \\
& \textbf{PseUI (UK)} &\textbf{67.33±0.06} &\textbf{68.79±0.01} & \textbf{72.73±0.09} & \textbf{0.33±0.008}& \textbf{0.69±0.01}\\
& {PSe-MA (rbf)} &{91.05}&{94.60}&{92.83} &{0.86}&{0.93}\\
& {PSe-MA (linear)} &{91.37}&{95.56}&{93.47} &{0.87}&{0.93}\\
& \textbf{PSe-MA (UK)} &\textbf{98.49±0.06}&\textbf{98.49±0.05}&\textbf{98.41±0.07} &\textbf{0.97±0.01}&\textbf{0.98±0.08}\\
\hline  
M994&iRNA-PseU (rbf)& 73.31& 64.83 &69.07& 0.38& 0.75\\
&iRNA-PseU (linear)&74.43 &63.31 &66.32 & 0.41&0.71 \\
&\textbf{iRNA-PseU (UK)}& \textbf{79.87±0.01}&\textbf{ 60.81±0.08}& \textbf{70.34±0.05}& \textbf{0.41±0.009}&\textbf{0.75±0.06}\\
& PseUI (rbf) &74.58& 66.31& 70.44& 0.41 &0.77\\
& PseUI (linear) &73.23  &68.92  &71.82  &0.51  &0.82\\
& \textbf{PseUI (UK)} &\textbf{75.78±0.009}& \textbf{68.76±0.07}& \textbf{73.22±0.04}& \textbf{0.64±0.01} &\textbf{0.79±0.09}\\
&{PSe-MA (rbf)}&  {77.69}&{77.38} & {77.35}&{0.56}&{0.79}\\
&{PSe-MA (linear)}&  {80.51}&{81.57} & {81.04}&{0.62}&{0.81}\\
&\textbf{PSe-MA (UK)} &\textbf{95.40±0.04}&\textbf{95.40±0.03} & \textbf{94.73±0.009}&\textbf{0.89±0.02}&\textbf{0.94±0.05}\\  
\hline
\end{tabular}}
 \caption{Jack-knife Cross-validation Results: iRNA-PseU and PseUI used RBF SVM on their data; our method PSe-MA used Utility Kernel on new biological features extracted in this work; the three feature vectors are made to run on three SVM kernels viz. RBF, linear and utility kernel. Utility Kernel outperformed the other methods with a Bonferroni Comparison (BC) value of 3.}
    \label{tab:jk_1}
\end{table}
\vspace*{-5mm}
\section{Results and discussions}
We evaluated the performance of different existing predictors with our proposed method, PSe-MA, and carefully tabulated the results in this section \footnote{ Tables of results, Code and Data Repository are at the \href{https://github.com/sourabh2711/Pseudouridine-predictor-PseMA}{github link}}. The parameters $k_0$ and $k_1$ of the proposed kernel are optimized by using the grid search technique. To investigate the performance of PSe-MA for $\psi$ site identification, the method is compared with the predictions from state-of-the-art ML models like XGBoost \cite{Chen2016XGBoostAS}, Random Forest \cite{Speybroeck2012ClassificationAR}, Decision Tree \cite{Quinlan1986InductionOD}, and Na\"ive Bayes \cite{Webb2005NotSN}. Further, we incorporated the jackknife, 5-fold cross-validation, and 66-33\% CV (results of 5-fold CV and 66-33\% CV in the supplementary file, section A, B) techniques to ensure meticulous performance validation across all the models. The brief description of the notations used is as follows. PSe-MA is the proposed method with new feature vectors (refer to section 3.1) which uses the proposed Utility kernel to predict $\psi$ sites. The other baselines - iRNA-PseU \cite{Chen2016iRNAPseUIR} and PseUI \cite{He2018PseUIPS}  used RBF kernel on their own feature vectors.  The derived features from each of the predictors are empirically evaluated on different kernels - Linear, RBF, and Utility Kernel. The cross-examination of features on these methods is presented to explain the in-depth analysis of the contributions of Utility Kernel on $\psi$ site identification.

A different set of experiments involves a comparison of XGBoost, Random Forest, Decision Tree, and Na\"ive Bayes with Utility Kernel on the newly proposed features (of PSe-MA). Table \ref{tab:jk_1}, 
compares the performance of features from iRNA-PseU, PseUI, and PSe-MA on RBF, linear, and Utility Kernel and, table \ref{tab:jk_2}, 
compares the performance of Utility Kernel with XGBoost, Random Forest, Decision Tree, and Naive Bayes on (only) PSe-MA features by using jackknife CV. 
The results of the 5-fold CV and 66-33\% CV are in the supplementary file sections A, and B; tables 1-4. While performing the experiments, the regularization parameter, C, and, the width parameter, $\gamma$ of the RBF kernel are optimized via grid search. Correspondingly, the $n\_estimators$ for Random forest and Decision Tree were taken to be the best-performing values. It is evident from the tables that PSe-MA (Utility Kernel) has outperformed the standard baselines (iRNA-PseU, PseUI) and the other competent ML models by a large margin for predicting the presence of $\psi$ site in the RNA samples. As per the results listed in table \ref{tab:jk_1}, PSe-MA is 32.2\% and 28.8\%  more accurate than iRNA-PseU and PseUI respectively on the H990 dataset, while it is 34.6\% and 32.6\% more accurate than iRNA-PseU and PseUI on S628 dataset and its 26.9\% and 25.5\% more accurate than iRNA-PseU and PseUI on M994 dataset for the task of $\psi$ site identification.\\
\noindent \textbf{Results of Generalization Settings:} Even under this novel setting, where the train and validation data are designed to be completely different (so as to completely overrule any possibility of leak validation), our method PSe-MA (Utility Kernel) shows SOTA results outperforming all existing baselines. The results are tabulated in Table 5 section B of the supplementary file where PSe-MA gave 95.40\% and 99.68\% accuracy in the two datasets thus surpassing the best accuracies from XGboost and Random Forest by a fair margin. It's important to note, however, that, the absolute performance is of lower priority. The ability of the Utility Kernel to adapt to completely unseen data and be able to generalize well is the pertinent point!

\begin{table}[]
    \centering
    \scalebox{0.65}{
 \begin{tabular}{ |p{1.5cm}|p{3cm}|p{2.0cm}|p{2.0cm}|p{2.0cm}|p{2.0cm}|p{1.8cm}|}
\hline
\multicolumn{7}{|c|}{Different ML models on the three training datasets using PSe-MA (Jackknife CV)} \\
\hline
Training datasets& Model used &Sensitivity&Specificity&Accuracy& MCC&AUC\\
\hline
H990 
& Gaussian NB &87.07&  89.70& 88.38& 0.77& 0.88\\
& Decision Trees &81.21& 78.59 & 79.90 &  0.60 & 0.80\\
& Random Forest &89.90 & 91.11 & 90.51& 0.81 & 0.91\\
& XGBoost &91.52& 91.31 & 91.41 & 0.83 & 0.91\\
& \textbf{PSe-MA (UK)} &\textbf{90.04±0.05}&\textbf{91.14±0.06}&\textbf{90.60±0.002} &\textbf{0.84±0.09}&\textbf{0.90±0.08}\\
   \hline
S628 
& Gaussian NB &93.29& 96.83& 95.06& 0.90&  0.95\\
& Decision Trees & 83.71 &  91.75 & 87.74 &0.76 & 0.88\\
& Random Forest &97.76 & 98.10 & 97.93 & 0.96 & 0.98\\
& XGBoost &97.12 &98.41 & 97.77 &  0.96 & 0.98\\
& \textbf{PSe-MA (UK)} &\textbf{98.49±0.02}&\textbf{98.49±0.01}&\textbf{98.41±0.08} &\textbf{0.97±0.006}&\textbf{0.98±0.09}\\ 
  \hline
M994
& Gaussian NB &90.25&  87.71& 88.98& 0.78 & 0.89\\
& Decision Trees & 84.32& 83.90 & 84.11 & 0.68 &0.84\\
& Random Forest &91.31 & 89.62 & 90.47 & 0.81 & 0.90\\
& XGBoost & 91.74& 91.53 & 91.63 & 0.83 & 0.92\\
&\textbf{PSe-MA (UK)}&  \textbf{95.40±0.08}&\textbf{95.40±0.01} & \textbf{94.73±0.05}&\textbf{0.89±0.04}&\textbf{0.94±0.06}\\  
\hline
\end{tabular}}
 \caption{Different ML models (Jackknife CV) compared with UK on our new set of features. The performance of PSe-MA is at par with XGBoost for H990 data, however, it outperformed the other models for the remaining two data sets with a Bonferroni Comparison (BC) value of 3.}
    \label{tab:jk_2}
\end{table}

\section{Conclusion}
According to the central dogma of molecular biology, RNA is the bridging molecule between the information storage molecule, DNA, and the information carrier molecule, protein. Besides the canonical A,G, C, and U bases, RNA contains approximately 120 modified bases and sugars. These modified bases and sugars play an essential role in the structural diversity and function of RNA. The most abundant modification found in RNA is pseudouridine, an isomerized form of uridine. Owing to the prohibitive costs and time associated with determining the $\psi$ sites in an RNA experimentally, computational prediction methods are often used. In this study, we present 'Pse-MA', a data-driven, utility kernel-based SVM model to predict the $\psi$ sites given the sequence information alone. PSe-MA has shown a significant increase in prediction accuracy (0.95, on average, compared to 0.72 of the state-of-the-art models). When validated using two unseen datasets, Pse-MA has achieved an accuracy of 97.54\% (average). Apart from being the best-performing prediction model among the set of existing baselines and accepted benchmark methods, our model, Pse-MA (SVM Utility Kernel) is able to generalize remarkably well on unseen test/validation data. Based on the current results, we are quite encouraged to apply this methodology to develop prediction methods for the other important functional modifications in RNA such as methylated Adenines and Cytidines. It would be worthwhile to explore if Deep learning-based feature extraction helps in further improvement of the prediction performance (preferably very close to $100\%$) given we are able to generate a sufficiently large corpus.

\newpage
\bibliography{main}
\bibliographystyle{plain}





\newpage

\appendix
\section*{Appendix}
\section{Results: PSe-MA compared with PseUI and iRNA-PseU model; 5-fold cross validation}
 In K-fold CV, the data is divided into K-folds. The $k-1$ folds are used for training and the $k^{th}$ fold is tested, and the process is repeated $k$ times, each time with a different subset for train and test (refer tables \ref{tab:5fcv} and \ref{tab:5fcv_all} ).

\begin{table}[h]
    \centering
    \scalebox{0.65}{
 \begin{tabular}{ |p{1.5cm}|p{3cm}|p{2.0cm}|p{2.0cm}|p{2.0cm}|p{2.0cm}|p{1.8cm}|}
\hline
\multicolumn{7}{|c|}{SVM Utility kernel against PseUI and iRNA-PseU models; with 5-fold CV} \\
\hline
Training datasets& Model used &Sensitivity&Specificity&Accuracy& MCC&AUC\\
\hline
H990  & iRNA-PseU (rbf)&63.21& 62.34& 61.32& 0.43&0.65\\
 & iRNA-PseU (linear)&65.32& 64.32& 63.89& 0.41&0.58\\
&  \textbf{iRNA-PseU (UK)}&\textbf{66.21±0.001} &\textbf{59.98±0.02} & \textbf{61.92±0.07}& \textbf{0.34±0.008} &\textbf{0.55±0.01}\\
& PseUI (rbf)&64.82& 63.25& 64.87& 0.56&0.68\\
& PseUI (linear)&65.78& 64.65& 66.66& 0.47&0.74\\
&\textbf{PseUI (UK)}& \textbf{70.76±0.09} & \textbf{69.54±0.07} & \textbf{71.33±0.08} &\textbf{0.46±0.05} &\textbf{ 0.50±0.04}\\
 & {PSe-MA (rbf)} &{70.30}&{81.62}&{75.96} &{0.52}&{0.76}\\ 
& {PSe-MA (linear)} &{63.03}&{76.97}&{70.00} &{0.41}&{0.70}\\ 
& \textbf{PSe-MA (UK)}& \textbf{89.52±0.01}&\textbf{89.52±0.031}&\textbf{89.49±0.02} &\textbf{0.78±0.06}&\textbf{0.89±0.071}\\ 
\hline
S628 &iRNA-PseU (rbf) &64.65 &64.33 &64.49 &0.29& 0.81\\
&iRNA-PseU (linear)&67.43& 67.83& 67.77& 0.43& 0.69\\
&\textbf{iRNA-PseU (UK)}&\textbf{68.55±0.06}& \textbf{69.33±0.09}& \textbf{68.61±0.04}& \textbf{0.51±0.05}& \textbf{ 0.69±0.02}\\
& PseUI (rbf) &62.10 &71.02 &66.56& 0.33& 0.69\\
& PseUI (linear) &64.32 &71.90 &65.65& 0.34& 0.61\\
& \textbf{PseUI (UK) } &\textbf{67.33±0.04} &\textbf{68.79±0.01} & \textbf{72.73±0.05} & \textbf{0.33±0.06}& \textbf{0.69±0.02}\\
& {PSe-MA (rbf)} &{90.10}&{94.92}&{92.51} &{0.85}&{0.93}\\
& {PSe-MA (linear)} &{69.01}&{93.02}&{81.05} &{0.81}&{0.64}\\ 
& \textbf{PSe-MA (UK)} &\textbf{97.69±0.07}&\textbf{97.69±0.04}&\textbf{97.61±0.01} &\textbf{0.95±0.02}&\textbf{0.97±0.05}\\ 
\hline

M994&iRNA-PseU (rbf)& 72.32& 61.32 &65.4& 0.32& 0.75\\
&iRNA-PseU (linear)& 73.31& 61.83 &63.07& 0.48& 0.65\\
&\textbf{iRNA-PseU (UK)}& \textbf{79.87±0.05}& \textbf{60.81±0.01}& \textbf{70.34±0.07}& \textbf{0.41±0.008}& \textbf{0.75±0.06}\\
& PseUI (rbf) &73.67& 64.33& 70.14& 0.53 &0.67\\
& PseUI (linear)&74.58& 66.31& 70.44& 0.41 &0.77\\
& \textbf{PseUI (UK) }&\textbf{75.43±0.08}&\textbf{ 67.73±0.03}& \textbf{71.32±0.06}& \textbf{0.39±0.02} &\textbf{0.79±0.01}\\
&{PSe-MA (rbf)}&  {78.62}&{77.34} & {77.97}&{0.56}&{0.78}\\
& {PSe-MA (linear)} &{75.44}&{74.80}&{77.92} &{0.56}&{0.78}\\ 
& \textbf{PSe-MA (UK)} & \textbf{91.45±0.007}&\textbf{91.45±0.01} & \textbf{91.42±0.09}&\textbf{0.83±0.04}&\textbf{0.91±0.02} \\  

\hline
\end{tabular}}
 \caption{5 fold Cross-validation Results-iRNA-PseU and PseUI used RBF SVM on their data; our method PSe-MA used Utility Kernel on new biological features extracted in this work; the three feature vectors are made to run on three SVM kernels viz. RBF, linear and utility kernel. Utility Kernel outperformed the other methods with a Bonferroni Comparison (BC) value of 3 on every dataset.}
    \label{tab:5fcv}
\end{table}








\begin{table}[h]
    \centering
    \scalebox{0.65}{
 \begin{tabular}{ |p{1.5cm}|p{3cm}|p{2.0cm}|p{2.0cm}|p{2.0cm}|p{2.0cm}|p{1.8cm}|}
\hline
\multicolumn{7}{|c|}{Different models (ML) on the three training datasets using PSe-MA} \\
\hline
Training datasets& Model used &Sensitivity&Specificity&Accuracy& MCC&AUC\\
\hline
H990 
& GausianNB & 88.41	&88.41	&88.48	&0.77	&0.88 \\
& Decision Trees &74.07	&74.07	&74.24	&0.48	&0.74 \\
& Random Forest &75.97	&75.97	&75.96	&0.52 &	0.75
\\
& XGBoost & 89.48&	89.48	&89.49	&0.78	&0.89 \\
& \textbf{PSe-MA (UK)} &\textbf{89.52±0.001 }&\textbf{89.52±0.004}&\textbf{89.49±0.02} &\textbf{0.78±0.03}&\textbf{0.89±0.07}\\
 \hline  
S628 
& GausianNB &94.78	& 94.78	&94.75	&0.89	&0.94 \\
& Decision Trees &85.59	&85.59	&85.51	&0.71	&0.85 \\
& Random Forest &83.1&	83.1&	83.13&	0.66&	0.83\\
& XGBoost & 97.04	&97.04	&96.98	&0.93	&0.97\\

& \textbf{PSe-MA (UK)} &\textbf{97.69±0.03}&\textbf{97.69±0.04}&\textbf{97.61±0.09} &\textbf{0.95±0.02}&\textbf{0.97±0.07}\\ 
\hline
  
M994
& GausianNB &88.93&	88.93&	88.88&	0.77&	0.88
\\
& Decision Trees & 80.66&	80.66&	80.62&	0.61&	0.8 \\
& Random Forest &77.84	&77.84	&77.75	&0.55	&0.77\\
& XGBoost & 90.88&	90.88&	90.00&	0.81&	0.9\\
&\textbf{PSe-MA (UK)}&  \textbf{91.45±0.01}&\textbf{91.45±0.03} & \textbf{91.42±0.01}&\textbf{0.83±0.10}&\textbf{0.91±0.08} 
\\  
\hline
\end{tabular}}
 \caption{Different ML models (5-fold CV) compared with our new set of features. Utility Kernel outperformed the other methods with a Bonferroni Comparison (BC) value of 3 on every dataset. }
    \label{tab:5fcv_all}
\end{table}
\section{Results: PSe-MA compared with PseUI and iRNA-PseU models; 66-33\% split}
The 66\%-33\% train-test split is the standard way of CV where 66\% of data is used for training and 33\% of data is used for reporting the performance of the model (refer tables \ref{tab:cv2/3} and \ref{tab:cv23_all}).
\begin{table}[]
    \centering
    \scalebox{0.70}{
 \begin{tabular}{ |p{1.5cm}|p{3cm}|p{2.0cm}|p{2.0cm}|p{2.0cm}|p{2.0cm}|p{1.8cm}|}
\hline
\multicolumn{7}{|c|}{PSe-MA compared with PseUI and iRNA-PseU models; 66-33\% split} \\
\hline
Training datasets& Training datasets &Sensitivity&Specificity&Accuracy& MCC&AUC\\
\hline
H990 
& iRNA-PseU (rbf) &62.41& 60.23& 62.44& 0.32&0.56\\
&  iRNA-PseU (linear) &60.32 &61.24 &63.43 &0.31 & 0.58\\
& \textbf{iRNA-PseU (UK)} & \textbf{62.34±0.03} & \textbf{58.65±0.02} &  \textbf{63.22±0.07} & \textbf{ 0.27±0.05} & \textbf{0.59±0.06}\\
& PseUI (rbf)&63.45& 65.64& 68.91& 0.38&0.64\\
& PseUI (linear) &68.93 & 66.80& 65.43& 0.32&0.52\\
& \textbf{PseUI (UK)} & \textbf{71.11±0.09} &  \textbf{68.34±0.03} &  \textbf{70.32±0.002} &  \textbf{0.44±0.04} &  \textbf{0.61±0.01}\\
& {PSe-MA (rbf)} &{79.81}&{76.21}&{77.32} &{0.42}&{0.71}\\ 
& {PSe-MA (linear)} &{77.78}&{80.61}&{79.20} &{0.79}&{0.58}\\ 
& \textbf{PSe-MA (UK)} &\textbf{90.18±0.03}&\textbf{90.18±0.01}&\textbf{90.2±0.09} &\textbf{0.8±0.04}&\textbf{0.9±0.009}\\
\hline
S628 
&iRNA-PseU (rbf) &65.45 &63.76 &65.76&0.38& 0.73\\
&  iRNA-PseU (linear) &62.45 & 64.44&62.32& 0.31&0.77\\
&\textbf{iRNA-PseU (UK)} & \textbf{68.55±0.07}&  \textbf{69.33±0.01}&  \textbf{68.61±0.06}& \textbf{ 0.51±0.08}&  \textbf{0.69±0.02}\\
& PseUI (rbf) &64.32 &75.02 &68.62& 0.54& 0.54\\
& PseUI (linear) &64.32& 72.31& 67.43& 0.23&0.64\\
& \textbf{PseUI (UK)} & \textbf{67.33±0.06} & \textbf{68.79±0.03} &  \textbf{72.73±0.01} &  \textbf{0.33±0.05}&  \textbf{0.69±0.02}\\
& {PSe-MA (rbf)} &{79.99}&{76.54}&{75.45} &{0.77}&{0.83}\\
& {PSe-MA (linear)} &{88.39}&{96.88}&{92.31} &{0.93}&{0.85}\\
& \textbf{PSe-MA (UK)} &\textbf{94.18±0.08}&\textbf{94.16±0.01}&\textbf{94.17±0.07} &\textbf{0.88±0.04}&\textbf{0.94±0.001}\\ 
\hline
M994
&iRNA-PseU (rbf)& 72.21& 65.23 &68.32& 0.42& 0.64\\
&  iRNA-PseU (linear) &72.45 &66.54 & 69.82&0.41 &0.76\\
&\textbf{iRNA-PseU (UK)}&  \textbf{79.87±0.08}&  \textbf{60.81±0.04}&  \textbf{70.34±0.06}&  \textbf{0.41±0.09}&  \textbf{0.75±0.007}\\
& PseUI (rbf)&74.21& 65.51& 71.23& 0.38 &0.67\\
& PseUI (linear) & 73.56& 67.83 & 71.21&0.45 &0.76\\
& \textbf{PseUI (UK)} & \textbf{76.58±0.01}&  \textbf{69.83±0.02}&  \textbf{71.32±0.04}& \textbf{ 0.52±0.07} & \textbf{0.65±0.05}\\
&{PSe-MA (rbf)}&  {78.82}&{79.32} & {75.43}&{0.61}&{0.61}\\
& {PSe-MA (linear)} &{77.50}&{80.92}&{79.17} &{0.79}&{0.58}\\ 
&\textbf{PSe-MA (UK)}& \textbf{91.0±0.07}&\textbf{91.0±0.009} & \textbf{91.0±0.01}&\textbf{82.0±0.05}&\textbf{91.0±0.09}\\  
\hline
\end{tabular}}
 \caption{ 66-33\% Cross-validation Results:iRNA-PseU and PseUI used RBF SVM on their data; our method PSe-MA used Utility Kernel on new biological features extracted in this work; the three feature vectors are made to run on three SVM kernels viz. RBF, linear, and utility kernel. Utility Kernel outperformed the other methods with a Bonferroni Comparison (BC) value of 3 on every dataset.}
    \label{tab:cv2/3}
\end{table}

\begin{table}[]
    \centering
    \scalebox{0.70}{
 \begin{tabular}{ |p{1.5cm}|p{3cm}|p{2.0cm}|p{2.0cm}|p{2.0cm}|p{2.0cm}|p{1.8cm}|}
\hline
\multicolumn{7}{|c|}{Different ML models on three training datasets (66-33 \% CV) using PSe-MA} \\
\hline
Training datasets& Training datasets &Sensitivity&Specificity&Accuracy& MCC&AUC\\
\hline
H990 
& GausianNB &88.9 &	88.9	&88.1	&0.77	&0.88
\\
& Decision Trees &77.95&	77.95&	77.98&	0.55&	0.77
\\
& Random Forest &78.56&	78.56&	78.59&	0.57&	0.78
\\
& XGBoost &90.1 &	90.1&	90.18&	0.8	&0.89
\\
& \textbf{PSe-MA (UK)} &\textbf{90.18±0.04}&\textbf{90.18±0.03}&\textbf{90.2±0.009} &\textbf{0.8±0.06}&\textbf{0.9±0.01}\\
\hline

S628 
& GausianNB &91.44	&91.44	&91.34	&0.82	&0.91 \\
& Decision Trees &80.57	&80.57	&80.28	&0.61&0.8  \\
& Random Forest & 84.6	&84.6	&84.61	&0.69&	0.84
\\
& XGBoost &93.83&	93.83&	93.75&	0.87&	0.93
\\
& \textbf{PSe-MA (UK)} &\textbf{94.18±0.02}&\textbf{94.16±0.04}&\textbf{94.17±0.07} &\textbf{0.88±0.09}&\textbf{0.94±0.08}\\ 
 \hline 
M994
& GausianNB & 89.1	&89.1	&89.1	&0.78	&0.89\\
& Decision Trees &81.71	&81.71	&81.41	&0.63	&0.81  \\
& Random Forest & 81.59	&81.59	&81.73	&0.63 &	0.81\\
& XGBoost & 89.51&	89.51&	89.42&	0.78	&0.89 \\
&\textbf{PSe-MA (UK)}&  \textbf{91.0±0.009}&\textbf{91.0±0.008} & \textbf{91.0±0.001}&\textbf{82.0±0.004}&\textbf{91.0±0.006}\\  
\hline
\end{tabular}}
 \caption{Different ML models (66-33\% CV) compared with our new set of features. Utility Kernel outperformed the other methods with a Bonferroni Comparison (BC) value of 3 on every dataset. }
    \label{tab:cv23_all}
\end{table}

\newpage
\section{Results on additional data}
In addition to the training datasets, Chen et al.provided two independent testing datasets for H. sapiens and S. cerevisiae, namely H200 and S200, but not for M. musculus. The results are in the table \ref{tab:5fcv_3}





\begin{table}[h]
    \centering
    \scalebox{0.65}{
 \begin{tabular}{ |p{1.5cm}|p{3cm}|p{2.0cm}|p{2.0cm}|p{2.0cm}|p{2.0cm}|p{1.8cm}|}
\hline
\multicolumn{7}{|c|}{5 fold CV with different ML models on testing datasets} \\
\hline
Testing datasets& Model used &Sensitivity&Specificity&Accuracy& MCC&AUC\\
\hline

H200 
& {Decision tress} &{77.82}&{78.92}&{77.79} &{0.74}&{0.84}\\
& {Random forrest} &{90.25}&{91.44}&{92.85} &{0.85}&{0.97}\\
& {XGBoost} &{91.72}&{90.25}&{93.28} &{0.87}&{0.82}\\
& {SVM rbf} &{74.75}&{77.14}&{75.97} &{0.85}&{0.67}\\
& {SVM linear} &{74.62}&{76.92}&{75.79} &{0.53}&{0.84}\\
& \textbf{SVM Utility} &\textbf{95.38±0.02}&\textbf{95.38±0.005}&\textbf{95.40±0.033} &\textbf{0.90±0.009}&\textbf{0.95±0.01}\\

\hline

S200
& {Decision tress} &{85.24}&{86.75}&{86.00} &{0.86}&{0.72}\\
& {Random forrest} &{99.04}&{99.04}&{99.04} &{0.97}&{0.98}\\
& {XGBoost} &{98.03}&{98.55}&{98.43} &{0.99}&{0.97}\\
&{SVM rbf}&  {96.65}&{96.14} & {96.38}&{0.99}&{0.93}\\
&{SVM linear}&  {89.61}&{89.64} & {89.63}&{0.96}&{0.79}\\
& \textbf{SVM Utility} &\textbf{99.69±0.07   }&\textbf{99.69±0.002}&\textbf{99.68±0.009} &\textbf{0.99±0.02}&\textbf{0.99±0.06}\\ 

\hline
\end{tabular}}
 \caption{5 fold Cross-validation Results with a comparison of different ML models on two of the testing datasets }
    \label{tab:5fcv_3}
\end{table}
\section{The sequence information of three Datasets- H.sapiens, S.cerevisiae, and M.musculus M.}
This study used three benchmark datasets for training: H990, S628, and M944. These are the same datasets used by iRNA-PseU and PseUI methods. These datasets are curated from the experimentally found $\psi$  sites from RMBase for H. sapiens, M. musculus, and S. cerevisiae. The negative dataset is made of RNA sequences that were experimentally confirmed to not have $\psi$ sites. In addition to the training datasets, Chen et al. \cite{Chen2016iRNAPseUIR} provided two independent testing datasets for H. sapiens and S. cerevisiae, namely H200 and S200, but not for M. musculus.
\begin{table}[h]
    \centering
    \scalebox{0.70}{
 \begin{tabular}{ |p{2cm}|p{2cm}|p{2cm}|p{2cm}|p{2cm}|}
\hline
Species & Name of the training dataset & Length of the RNA sequence & Number of positive samples& Number of negative samples\\
\hline
\textit{H.sapiens}& H-990 & 21 &495 & 495\\
\hline
\textit{S.cerevisiae}& S-628 & 31 &314 & 314\\
\hline
\textit{M.musculus}& M-994 & 21 & 472 & 472\\
\hline
\end{tabular}}
\caption{The sequence information of the three datasets}
\label{tab:desc}
\end{table}
\begin{table}[h]
    \centering
    \scalebox{0.70}{
 \begin{tabular}{ |p{2cm}|p{2cm}|p{2cm}|p{2cm}|p{2cm}|}
\hline
Species & Name of the training dataset & Length of the RNA sequence & Number of positive samples& Number of negative samples\\
\hline
\textit{H.sapiens}& H-200 & 21 &100 & 200\\
\hline
\textit{S.cerevisiae}& S-200 & 31 &100 & 200\\
\hline
\end{tabular}}
\caption{Additional Data - The sequence information of the two testing datasets}
\label{tab:desc}
\end{table}

\end{document}